\documentstyle[erice99,12pt,epsfig,graphics]{article}
\input epsf

\newcommand{\be}[1]{\begin{equation} \label{(#1)}}
\newcommand{\ee}{\end{equation}}
\newcommand{\ba}[1]{\begin{eqnarray} \label{(#1)}}
\newcommand{\ea}{\end{eqnarray}}

\begin{document}
\vspace{-1cm}
\hspace{12cm} JLAB-PHY-99-28
\title{\LARGE Recent Results from Jefferson Lab} 
\author{\large Volker D. Burkert\footnote[2]{e-mail: burkert@jlab.org}}
\address{Jefferson Lab, Newport News, Va. 23606, USA}
%
\maketitle

\abstracts{Recent results on studies of the structure of nucleons
and nuclei in the regime of strong interaction QCD are discussed. 
Use of high current polarized electron beams, polarized targets, and 
recoil polarimeters, in conjunction with modern spectrometers and detector 
instrumentation allow much more detailed studies of nucleon and nuclear 
structure than has been possible in the past. The CEBAF accelerator 
at Jefferson Lab was build to study the internal structure of hadrons in 
a regime where confinement is important and strong interaction QCD is 
the relevant theory. I discuss how the first experiments already 
make significant contributions towards an improved understanding of
hadronic structure.}

\section{Introduction}
                                                       
Electromagnetic production of hadrons may be characterized according to 
distance and time scales (or momentum and energy transfer) 
probed in the interaction. This is illustrated with the 
three regions in Figure 1. For simplicity I have omitted the time scale. 
At large distances mesons and nucleons are the 
relevant degrees of freedom. Due to the limited spatial resolution 
of the probe we study peripheral properties of nucleon and nuclei 
near threshold for pion production. Chiral perturbation theory describes many 
of these processes and has a direct link to QCD via (broken) chiral symmetry. 
 At short distances (and short time scales), the coupling involves elementary quark and 
gluon fields, governed  by perturbative QCD, and we map out parton distributions in the nucleon.  
At intermediate distances, quarks and gluons are 
relevant, however,  confinement is important, and they appear as constituent 
quarks and glue. We study interactions between these constituents via their excitation
 spectra and wave functions. This is the region where the connection to the fundamentals 
of QCD remains poorly established, and where JLab experiments currently have their 
biggest impact. This is the region I will be focusing on in this lecture. 
These regions are not strictly separated from each other but 
overlap, and the hope is that due to this overlap hadron structures 
may eventually
be described in a more unified approach based on fundamental theories.
Because the electro-magnetic and electro-weak probes are well understood, 
they are best suited to
provide the data for such an endeavor. I will discuss recent data
on studies of the intrinsic nucleon structure, and results from light nuclear targets, 
$^2H$ and $^3He$.

\begin{figure}[htbp]
\epsfysize=18.0truecm
\epsfbox{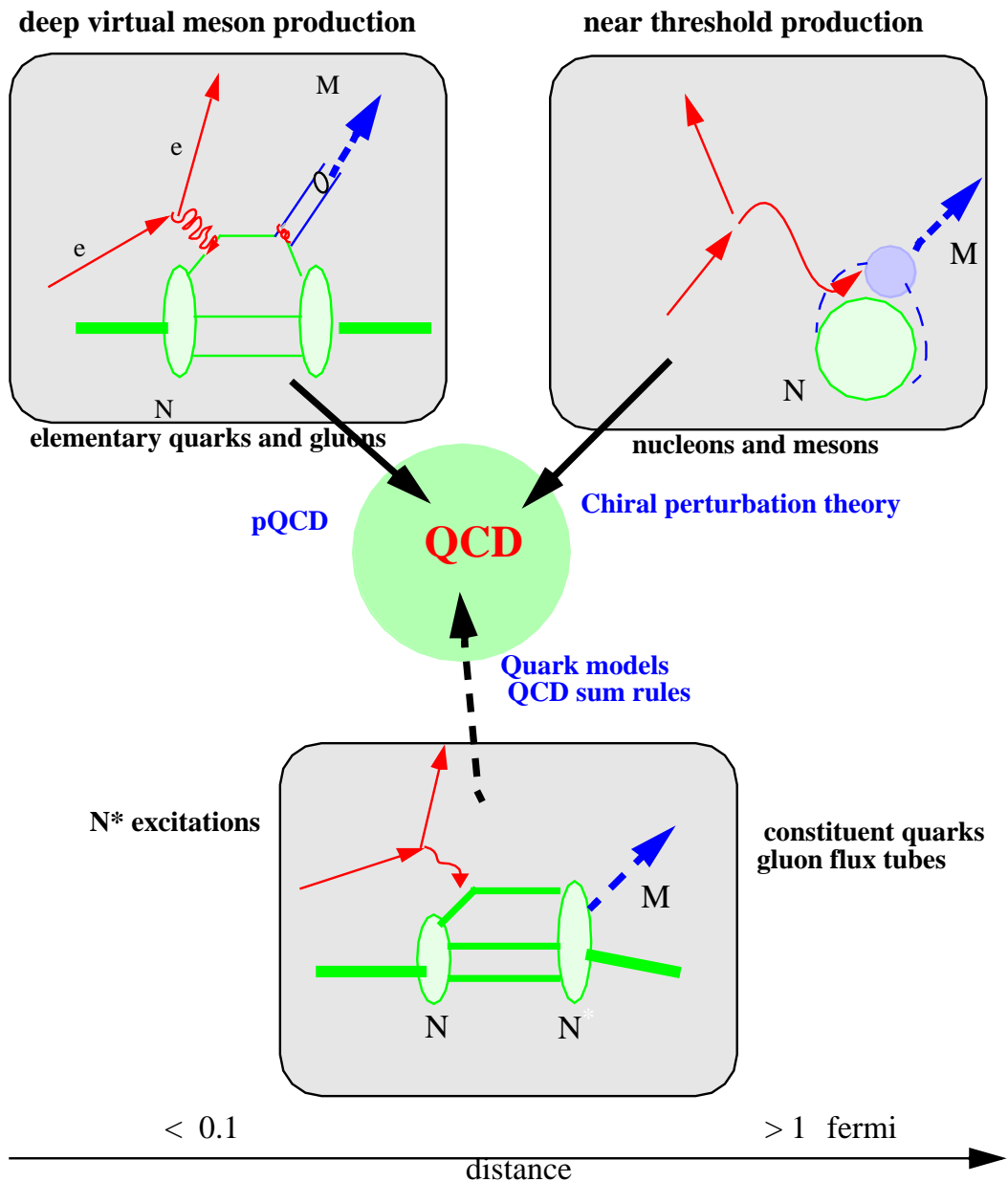}

\hsize=15cm
\caption{\small Exclusive meson electroproduction. A subdivision in
distance scales is used to illustrate three kinematic regions and their 
respective (effective) degrees of freedom.}

\end{figure}

\vspace{0.5cm}

\subsection{\bf Structure of the Nucleon at Intermediate Distances - Open Problems}

QCD has not been solved for processes at intermediate distance scales. A direct 
consequence is that the internal structure of nucleons is generally 
poorly known in this regime. On the other hand, theorists are often 
not challenged due to the lack of high quality data in many areas. 
The following are areas where the lack of high quality data is most noticeable, and
where JLab experiments have already contributed, or are expected to contribute
significantly in the future:

\begin{itemize}

\item{The electric form factors of the nucleon $G_{En}$, $G_{Ep}$ are poorly known, 
especially for the neutron, but also for the proton. This means that the 
charge distribution of the basic building blocks of the most common form of
matter in the universe is virtually unknown.} 

\item{We do not know what role strange quarks play in the wave function of ordinary 
matter. }

\item{The nucleon spin structure has been explored for more than two decades 
at high energies in laboratories such as CERN and SLAC. The confinement regime and
the transition to the deep inelastic regime have not been explored at all.}  

\item{To understand the ground state nucleon we need to understand
the full excitation spectrum as well as the continuum. 
Few transitions to excited states have been studied well, and many
states are missing from the spectrum as predicted by our most accepted models.} 

\item {The role of the glue in the baryon excitation spectrum is completely unknown, 
although gluonic excitations of the nucleon are expected to be produced 
copiously \cite{isgur}.}  
 
\item {The long-known connection between the deep inelastic regime and the 
regime of confinement (local duality) \cite{blogil} 
remained virtually unexplored for decades.}

\end{itemize}

{\sl Carrying out an experimental program that will address these questions 
has become feasible due to the 
availability of CW electron accelerators, modern detector instrumentation with 
high speed data transfer techniques, the routine availability of spin
polarization in beam and targets and recoil polarimetry.}

\vspace{0.3cm}

The main contributor to this field is now the CEBAF accelerator at Jefferson Lab 
in Newport News, Virginia, USA.
A maximum energy of ~6 GeV is currently available. The three
 experimental halls (A, B, C) can receive polarized beam simultaneously, 
with different but correlated beam energies, or with the same beam energies. 
This allows a diverse 
physics program to be carried out in a very efficient way.

\section{Structure of the Ground State Nucleon}

\subsection{\bf Charge and current distribution} 

The nucleon ground state has been studied for decades in elastic electron-nucleon 
scattering. It
probes the charge and current distribution in the nucleon 
in terms of the 
electric ($G^{\gamma}_E$)and magnetic ($G^{\gamma}_M$) form factors. The superscript
$\gamma$ or $Z$ are used to describe the electromagnetic and weak form factor, 
respectively.
Early experiments from Bonn, DESY, and CEA showed a violation of the so-called 
"scaling law", which may be interpreted that the spatial distribution  of 
charge and magnetization are not the same, and the 
corresponding form factors have different $Q^2$ dependencies. The data showed a 
downward trend for the ratio $R^{\gamma}_{EM} = G^{\gamma}_E/G^{\gamma}_M$ as a function of $Q^2$.  
Adding the older and newer SLAC data sets confuses the picture greatly (Figure 2). 
Part of the  data are incompatible with the other data sets. They also do not 
show the same general trend. 
Reliable data were urgently needed to clarify the experimental situation and to constrain 
theoretical developments. 
 
Reliable data for the electric form factors at high $Q^2$ can be obtained
 using double polarization measurements, 
and the first experiments of this type have now produced results. For a specific kinematics
where the proton polarization is measured in the electron scattering plane, but
transverse to the 
virtual photon, the 
double polarization asymmetry is given by:
$$A_{\vec e\vec p} = {k_1{R^{\gamma}_{EM}} \over {k_2(R^{\gamma}_{EM})^2 + k_3}}~~,$$ where 
the $k_i$ are kinematic quantities. 
Since the ratio $R^{\gamma}_{EM}$ is accessed directly this experiment has 
smaller systematic uncertainties 
than previous experiments at high $Q^2$ (Figure 2). They confirm the trend of the 
early data, improve the accuracy at high $Q^2$ significantly, and extend the $Q^2$ range. 
The data illustrate beautifully the power of polarization in electromagnetic 
interactions. The experiment will be continued to higher momentum transfer in 
the year 2000. Other experiments\cite{day,madey} will measure the same quantity on the 
neutron from a deuterium target using a similar techniques.

A precision measurement of neutron magnetic form factor will be carried out with 
CLAS using the 
neutron to proton ratio measured simultaneously \cite{brooks}. This experiment 
will use the reaction $ep \rightarrow en\pi^+$ for an in-situ calibration of 
the neutron counter detection efficiency.

\subsection{\bf Strangeness Structure of the Nucleon}

From the analysis of deep inelastic polarized structure function experiments we know 
that the strange quark sea is polarized, 
and  contributes at the 5 - 10\% level to the nucleon spin.  
Then one may ask what are the strange quark contributions to the nucleon 
ground state wave function and their corresponding form factors? 
The flavor-neutral photon coupling does not distinguish s-quarks from u- 
or d-quarks. However, the tiny contribution of the $Z^o$ is parity violating, 
and allows measurement of the strangeness contribution. 
The effect is measurable due to the interference with the 
single photon graph. The asymmetry $$A_{\vec ep} = {G_FQ^2\over \sqrt{2}\pi\alpha}
{{\epsilon G_E^{\gamma} G_E^Z + \tau G_M^{\gamma}G_M^Z - {1 \over 2}
(1-4sin^2\theta_W)KG^{\gamma}_MG^Z_A} \over 
{\epsilon (G_E^{\gamma})^2 + \tau (G_M^{\gamma})^2}} $$ 
\noindent
in polarized electron scattering contains 
combinations of electromagnetic and weak form factors. 
The term containing the axial form factor $G^Z_A$  is suppressed due to 
the factor $(1-sin^2\theta_W$), and gives small corrections. 
The weak form factors can be 
expressed in terms of the $G^{\gamma}$ and the strangeness form factors ($G^s$). 
For example, the weak electric form factor can be written:
$$ G_E^Z = ({1\over 4} - sin^2\theta_W)G^{\gamma}_{Ep} - {1\over 4} 
(G^{\gamma}_{En} + G^s_E)$$
\noindent
The same relation holds for the magnetic form factors. 
The $G^s$ form factors can be measured since the $G^{\gamma}$ are known.
The elastic $\vec e p$ results of the JLAB HAPPEX experiment measured at
$Q^2 = 0.47 GeV^2$, show that strangeness contributions are small
when measured in a combination of $G^s_E$ and $G^s_M$ \cite{happex1}:

$$ G^s_E + 0.4G^s_M = 0.023 \pm 0.034 (stat) \pm 0.022 (syst) \pm 0.026 (G^n_E)$$
At least a factor of two smaller statistical error will 
be obtained when the 1999 data are included in the analysis. The error is then
dominated by the uncertainties in the neutron electromagnetic form factor, 
especially $G_{En}$! 
New measurements of $G^{\gamma}_{En}$ and 
$G^{\gamma}_{Mn}$ should remedy this situation \cite{day,madey,brooks}.

\section{The Nucleon Spin Structure - from Small to Large Distances} 

The internal spin structure of the nucleon has been of central interest ever since the 
EMC experiment 
found that at small distances the quarks carry only a fraction of the
nucleon spin. 
Going from small to large distances the quarks get dressed with gluons and $q\bar q$ pairs 
and acquire more and more of the nucleon spin. How is this process evolving with the distance scale? 
At the two extreme kinematic regions we have two fundamental sum rules: the 
Bjorken sum rule (Bj-SR) which holds in the asymptotic 
limit, and is usually written for the proton-neutron difference as 
$$ \Gamma_1^{pn} = \int{g_1(x)dx} = {g_A \over 6}~~.$$ At the finite $Q^2$ where 
experiments are performed, 
QCD corrections have been calculated, and there is good agreement between theory and experiment at 
$Q^2> 2~ GeV^2$.
At the other end, at $Q^2 = 0$, the Gerasimov Drell-Hearn sum rule (GDH-SR) is expected to hold: 
$$I_{GDH} = {M^2\over 8\pi^2\alpha}\int {{\sigma_{1/2}(\nu)-\sigma_{3/2}(\nu)}\over \nu}d\nu = 
-{1\over 4}\kappa^2 ~~ .$$
The integral for the difference in helicity 1/2 and helicity 3/2 total 
absorption cross sections 
 is taken over the entire inelastic energy regime. The quantity $\kappa$ is the anomalous 
magnetic moment of the target.

\begin{figure}[htbp]
\begin{minipage}{0.45\textwidth}
\epsfysize=7.0cm
\epsfxsize=7.5cm
\epsfbox{ 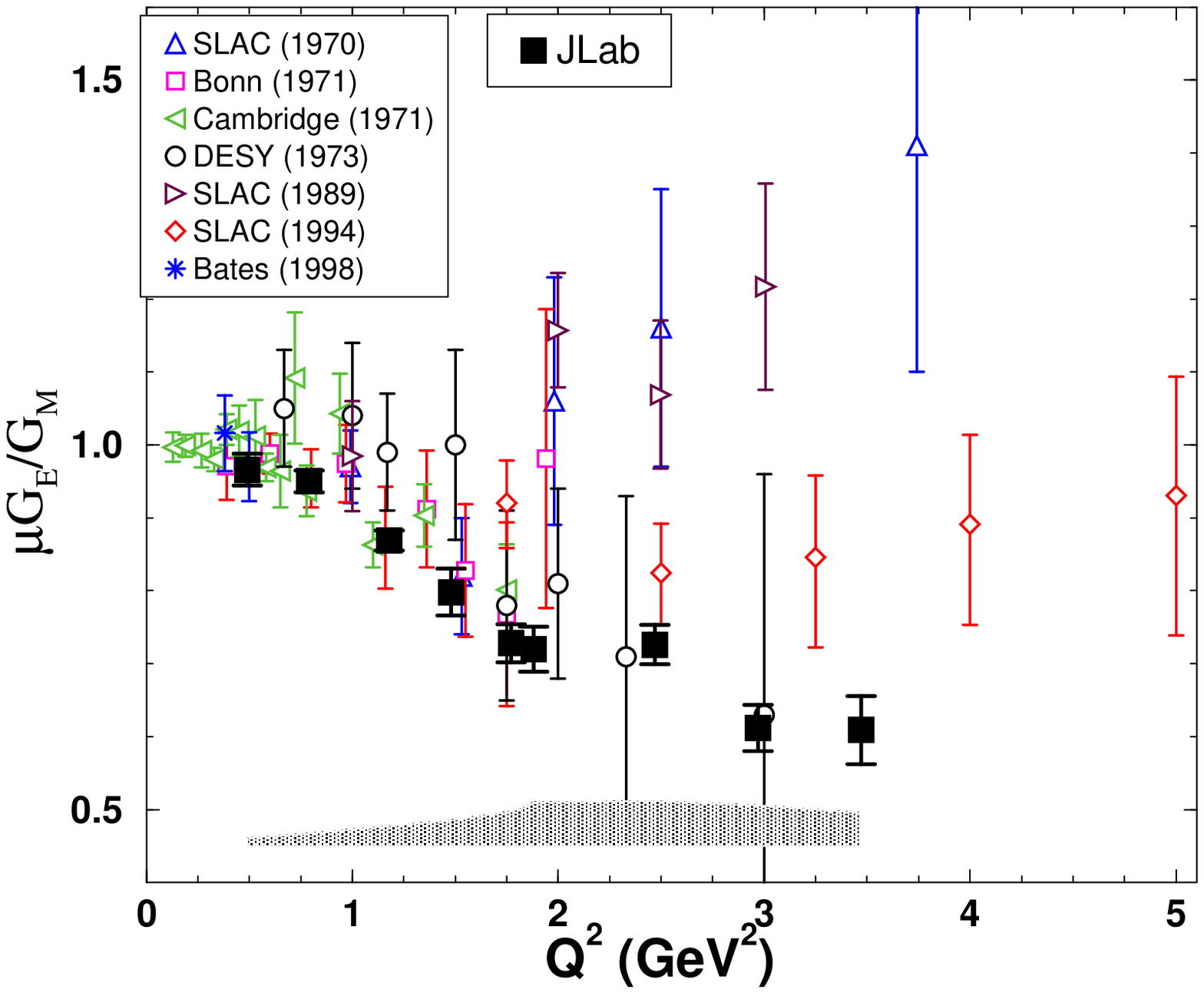}
\hsize=7.5cm
\caption{\small Results for the ratio $R^{\gamma}_{EM}$ of electric 
and magnetic form factors of the 
proton. The full squares are the results from JLAB obtained with the 
double polarization techniques \cite{perdrisat} }
\end{minipage}
\begin{minipage}{0.45\textwidth}
\epsfysize=9.0cm
\epsfxsize=7.5cm
\epsfbox{ 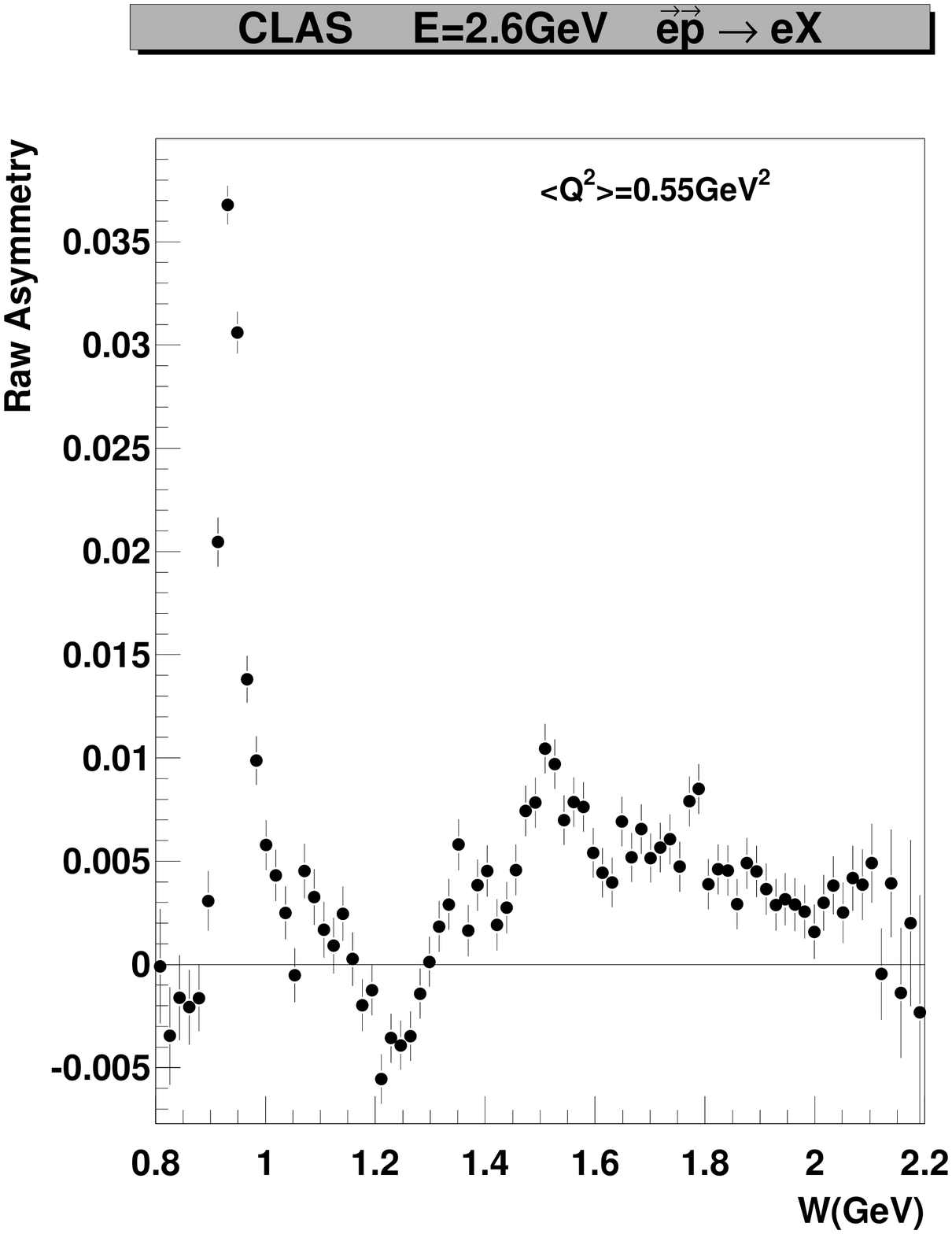}
\hsize=7.5cm
\caption{\small Raw asymmetry measured in inclusive $\vec e N{\vec H}_3 \rightarrow e X$
scattering at JLAB.}
\end{minipage}
\end{figure}

One important connection between these regions is given by the constraint due to the GDH-SR 
- it defines the slope of the Bjorken integral ($\Gamma_1^{pn}(Q^2) = \int g_1^{pn}(x,Q^2)dx$) at $Q^2=0$: 
$$ I^{pn}_{GDH}(Q^2 \rightarrow 0) = 2{M^2\over Q^2}  \Gamma_1^{pn} (Q^2 \rightarrow 0) $$ 
\noindent 
Phenomenological models have been proposed to extend the GDH integral for the 
proton and neutron to finite $Q^2$ and connect it to the deep inelastic regime 
\cite{buriof1,buriof2,soffer}. The data at low $Q^2$ \cite{e143} are in 
good agreement with the 
predictions if nucleon resonances are taken into account explicitly \cite{buriof2} 
(Figure 4).

\noindent
An interesting question is whether we can go beyond models and describe the 
transition from 
the Bj-SR to the  GDH-SR for the proton-neutron difference within the framework of 
fundamental theory, i.e. QCD. 
 For the proton and neutron, the GDH-SR is nearly saturated by low-lying 
 resonances \cite{burkert1,ma} with
the largest contributions coming from the excitation of the $\Delta(1232)$. 
The latter contribution is absent in  
the p-n difference. Other resonance contributions are reduced as well and 
the $Q^2$ evolution may take on a smooth transition to the Bj-SR regime. 
A crucial question in this connection is: how low in $Q^2$ the Bj-SR can be evolved 
using the modern techniques of higher order QCD expansion?
Recent estimates  \cite{ji1} suggest these techniques may be valid as low 
as $Q^2 = 0.5$ GeV$^2$. 
At the other end, at $Q^2=0$, where hadrons are the relevant degrees of freedom,
chiral perturbation theory may be applicable at very small $Q^2$, 
and may allow evolution of the GDH-SR to finite $Q^2$. 
Significant theoretical efforts are needed to bridge the remaining gap, 
perhaps utilizing lattice QCD. 
These efforts are of utmost importance since {\it it would mark the first time that 
hadronic structure is described by fundamental theory in the entire kinematic
regime, from small to large distances!}

\noindent
Experiments have been carried out at JLAB on $NH_3$ \cite{bucramin}, $ND_3$\cite{kuhtai}, 
and $^3He$ \cite{jpchen} 
targets to extract the $Q^2$ evolution of the GDH integral for protons and neutrons
in the low $Q^2$ range $Q^2 = 0.1 - 2.0~GeV^2$ and from the elastic to the deep 
inelastic regime. Currently, only two data points with large errors exist for 
$Q^2 < 2~ GeV^2$. Because of the current limitations in machine energy to 6 GeV, some 
extrapolation will be needed to determine the full integral, especially at the 
larger $Q^2$ values. The deep inelastic contributions to the GDH integral have been 
measured for $Q^2$ above 1.3 $GeV^2$ \cite{hermes}. 
First results from the JLAB experiments are expected in the year 2000. Figure 3 shows 
an uncorrected asymmetry from an experiment on polarized $NH_3$. The positive elastic 
asymmetry, the negative asymmetry in the $\Delta$ 
region, and the changeover back to a positive asymmetry for higher mass resonances and 
the high energy continuum are evident.

\begin{figure}[htbp]
\epsfysize=16.0truecm
\epsfxsize=14.0truecm
\begin{center}
\epsfbox{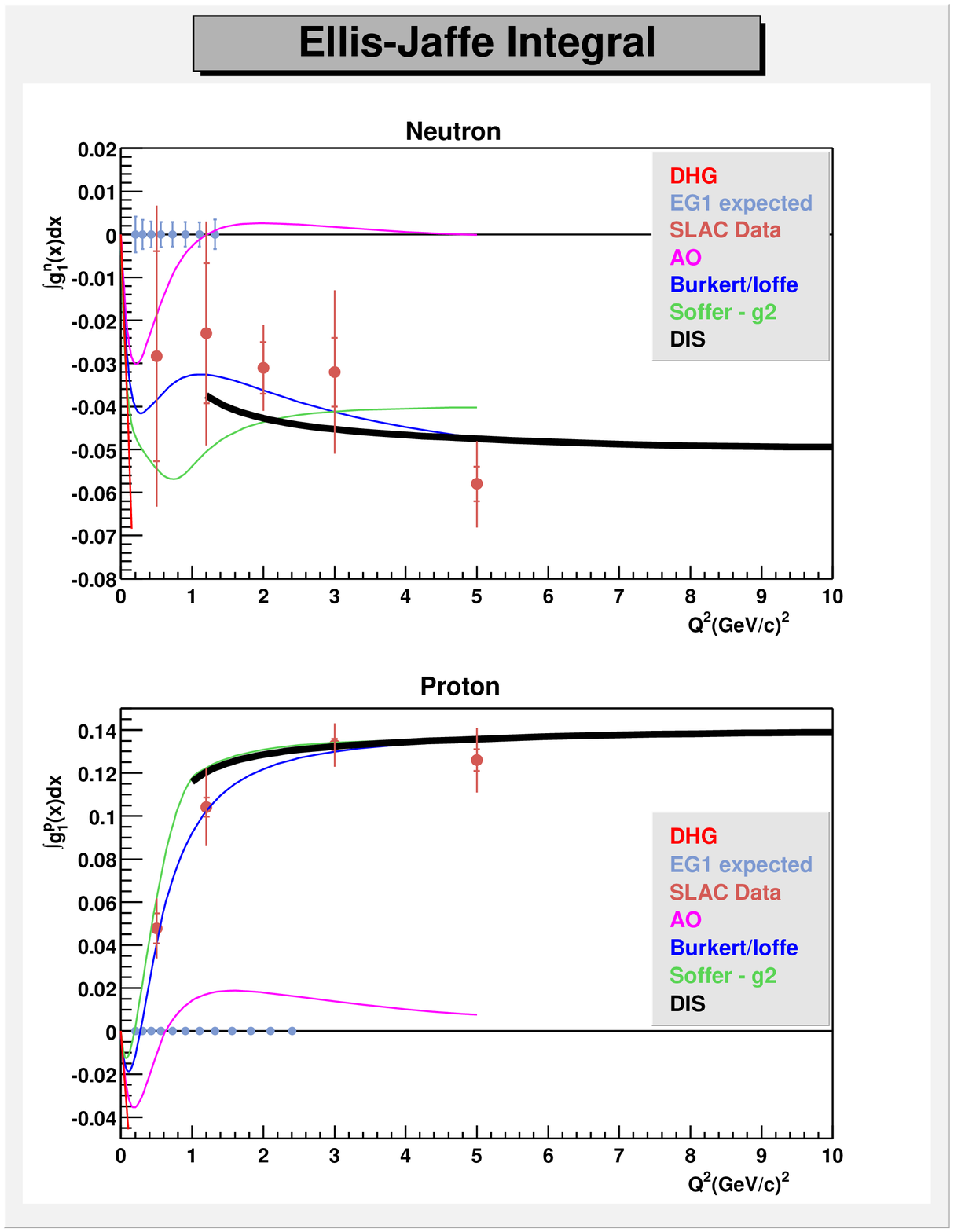}
\hsize=14cm
\vspace{1cm}
\caption{\small The first moment $\Gamma_1(Q^2)$ of the polarized structure function 
$g_1(x,Q^2)$. Model predictions from \cite{soffer,buriof2}. The curve labelled AO contains
s-channel resonance contributions only\cite{burkert1}. The straight line 
near $Q^2=0$ is the slope given by the GDH sum rule constraint. The points along the 
horizontal axis indicate the expected statistical errors for the measured portion 
of the integral on the proton $NH_3$ and neutron $ND_3$.}
\end{center}
\end{figure}

\section{Excitation of Baryon Resonances}

A large effort is being extended to the study of excited states of the nucleon. 
The  transition form factors contain information 
on the spin structure of the transition and the wave function of the 
excited state. We test predictions of baryon structure models and strong interaction 
QCD.
Another aspect is the search for, so far, unobserved states which 
are missing from the spectrum but are predicted 
by the QCD inspired quark model \cite{isgur2}. Also, are there other than $|Q^3>$ states? 
Gluonic excitations of the nucleon, i.e. $|Q^3G>$ states
may be be copious /cite{isgur}, and some resonances may be ``molecules'' of baryons and mesons 
$|Q^3Q\bar Q>$.
Search for at least some of these states is important to clarify the intrinsic 
quark-gluon structure of baryons and 
the role played by the glue and mesons in hadron spectroscopy and structure. Electroproduction
is an important tool in these studies as it probes the internal 
structure of hadronic systems.
The scope of the $N^*$ program\cite{nstar,ripani} at JLAB is to 
measure many of the possible decay channels of resonances in a large kinematic
range.

\subsection{\bf The $\gamma N \Delta$ transition.}

The lowest excitation of the nucleon is the $\Delta(1232)$ ground state. The 
electromagnetic excitation is due dominantly to a quark spin flip corresponding
to a magnetic dipole transition. 
The interest today is in measuring the small electric and scalar quadrupole
transitions which are predicted to be sensitive to possible deformation of the nucleon 
or the $\Delta(1232)$ \cite{delta}.
Contributions at the few percent level may come from the pion cloud at 
large distances, and gluon 
exchange at small distances. 
An intriguing prediction is that in the hard scattering limit the 
electric quadrupole contribution should be equal in strength to the 
magnetic dipole contribution \cite{carlson}. An analysis \cite{burelou} of earlier DESY data
found small nonzero values for the ratio $E_{1+}/M_{1+}$ at $Q^2 = 3.2GeV^2$, 
showing that the
asymptotic QCD prediction is far away from the data.    
 
An experiment at JLAB Hall C \cite{frolov} measured  $p\pi^o$ production in the 
$\Delta(1232)$ 
region at high momentum transfer, and found values for
$|E_{1+}/M_{1+}| < 5 \%$ up to $Q^2 = 4~GeV^2$.
Analysis of new data from CLAS indicate negative values at small $Q^2$ 
with a trend towards positive values at higher $Q^2$. Results should be
available in 2000.

\subsection {\bf Higher mass resonances}

The inclusive spectrum shows only 3 or 4 enhancements, however 
more than 20 states are known in the mass region up to 2 GeV. 
 By measuring the electromagnetic transition of many of these 
 states we can study symmetry properties 
 between excited states and obtain a 
 more complete picture of the nucleon structure. 
For example, in the single-quark-transition model only one quark
participates in the interaction. It predicts transition amplitudes 
for a large number of states based on a few measured amplitudes \cite{hey}.
The current situation is shown in Figure 5, where the SQTM amplitudes for 
the transition to the $L_{3q}=1$ $SU(6)\otimes O(3)$ multiplet have been extracted from
the measured amplitudes for $S_{11}(1535)$, and $D_{13}(1520)$. Predictions for other
states belonging to the same multiplet are shown in the other panels. The lack of accurate 
data for most other resonances prevents a sensitive test of even the simple algebraic SQTM.

\begin{figure}[htbp]
\epsfysize=18.0truecm
\hoffset=3.5truecm
\epsfbox{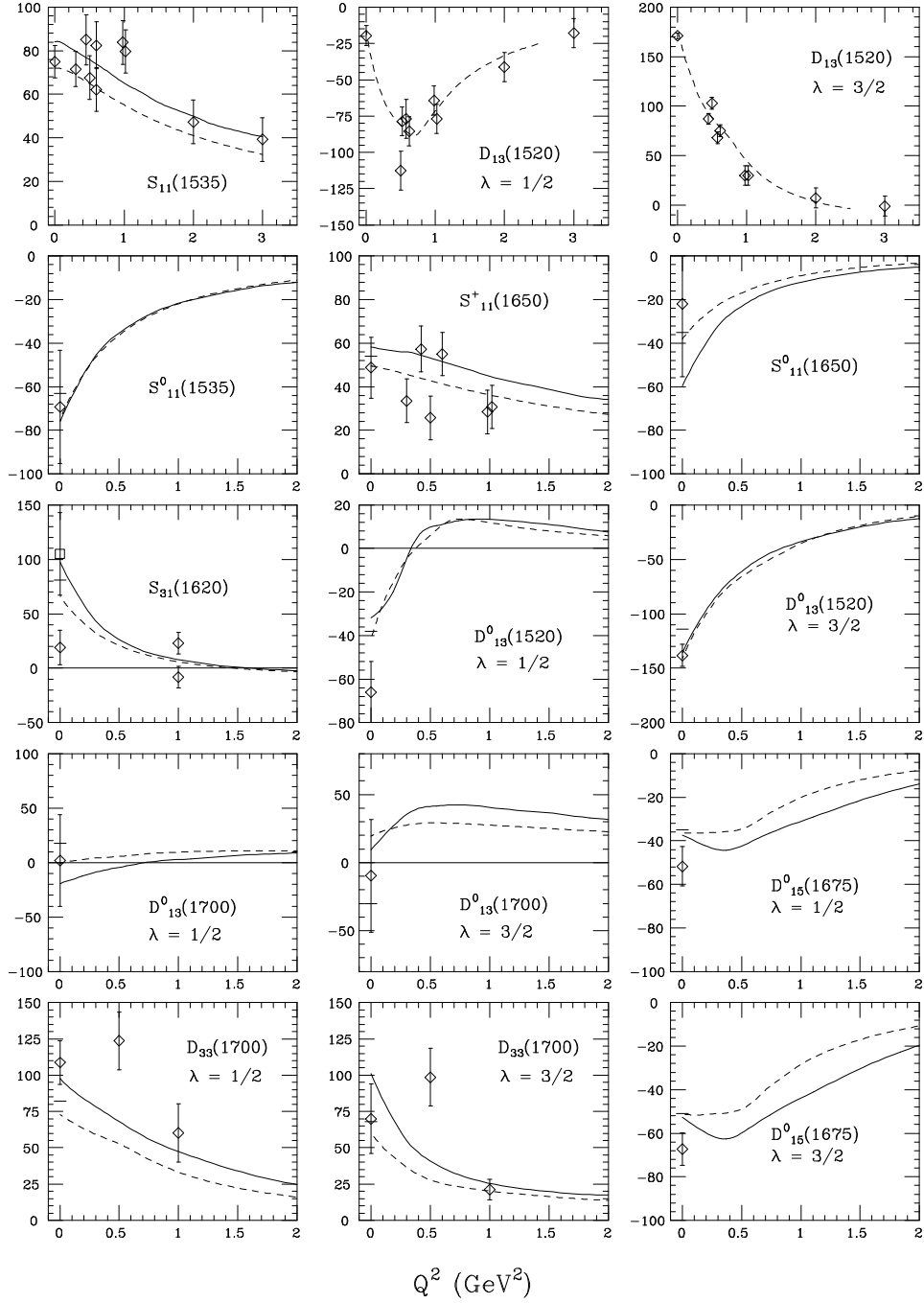}
\hsize=14cm
\caption{\small Single Quark Transition Model predictions for states 
belonging to the $SU(6) \otimes O(3)$ multiplet, discussed in the text.}
\end{figure}

The goal of the N* program at JLAB with the CLAS detector is to 
provide data in the entire resonance region, by measuring many
channels in a large kinematic range, including many polarization observables. 
The yields of several channels recorded simultaneously are shown in Figure 6
and Figure 7. Resonance excitations seem to be 
These yields illustrate how the various channels have different sensitivity 
to various resonance excitations. For example, the $\Delta^{++}\pi^-$ channel 
clearly shows resonance excitation near 1720 MeV while single pion 
production is more sensitive to a resonance near 1680 MeV 
\cite{ripani}. The $p\omega$ channel shows resonance excitation near threshold, 
similar to the $p\eta$ channel. No resonance has been observed in this channel 
so far. For the first time  $n\pi^+$ electroproduction has been measured 
throughout the resonance region, and in a large angle and $Q^2$ range.

\begin{figure}[htbp]
\begin{minipage}{0.45\textwidth}
\epsfysize=8.5truecm
\epsfxsize=8.0cm
\epsfbox{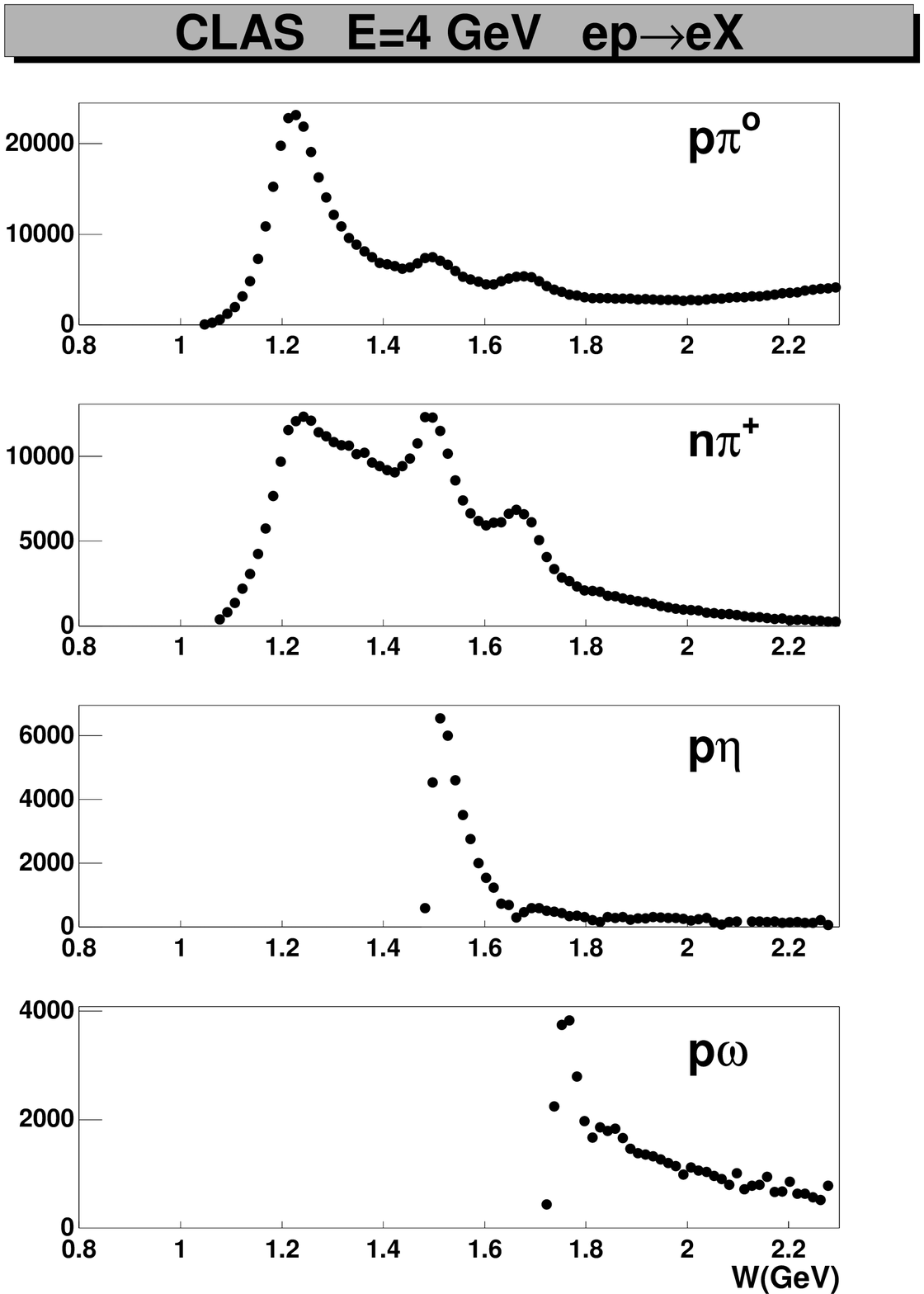}
\hsize=7.5cm
\caption{\small Yields for various channels measured with CLAS at JLAB. 
The statistical error bars are 
smaller than the data points.}
\end{minipage}
\begin{minipage}{0.45\textwidth}
\epsfysize=8.5truecm
\epsfxsize=8.0cm
\epsfbox{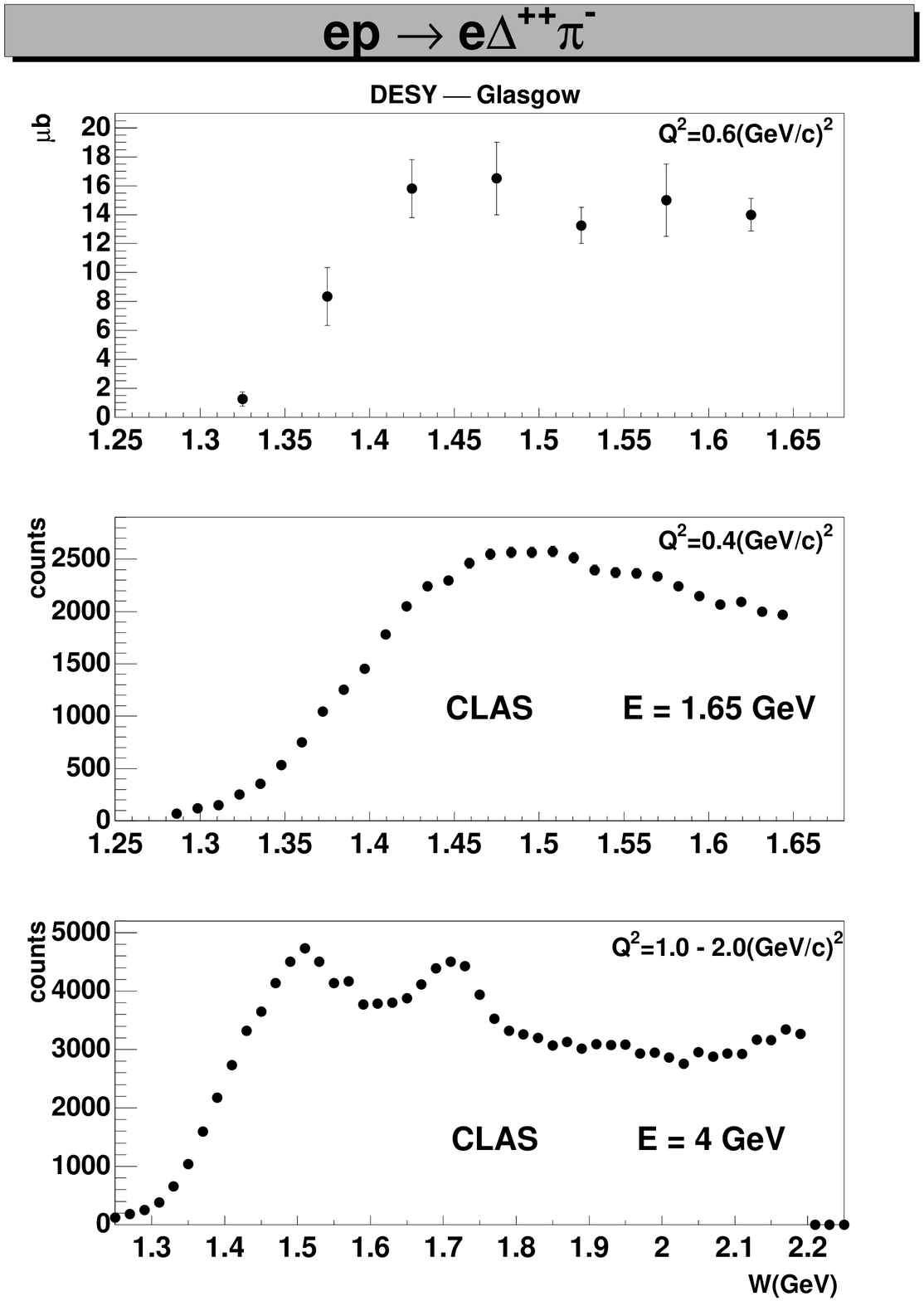}
\hspace{1cm}
\hsize=7.5cm
\caption{\small Yields for the channel $\Delta^{++}\pi^-$ measured with CLAS at 
different $Q^2$ compared to previous data from DESY.}
\end{minipage}
\end{figure}

Figure 7 illustrates  the vast improvement in data volume for the 
$\Delta^{++}\pi^-$ channel. 
The top panel shows DESY data taken more than 20 years ago.  The other 
two panels show samples of the data taken so far with CLAS. At higher $Q^2$, 
resonance structures, not seen before in this channel are revealed.

\subsection {\bf Missing quark model states} 

These are states predicted in the $|Q^3>$ model to populate the 
mass region around 2 GeV.  However, they have
not been seen in $\pi N$ elastic scattering, our main source of information 
on the nucleon excitation spectrum. 

How do we search for these states?
Channels which are predicted to couple strongly to these states are
$N(\rho, \omega)$ or $\Delta\pi$. 
Some may also couple to $KY$ or $p\eta^{\prime}$ \cite{capstick}. 

Figure 8 shows preliminary data from CLAS in $\omega$ production 
on protons. The 
process is expected to be dominated by diffraction-like $\pi^o$ 
exchange with strong peaking at forward $\omega$ angles, or low t,  
and a monotonic fall-off at large t. 
The data show clear deviations 
from the smooth fall-off for the W range near 1.9 GeV, where some of the 
``missing'' resonances are predicted, in comparison with the high W region. 

Although indications for resonance production are strong, analysis of 
more data and a full partial wave study are needed before 
definite conclusions may be drawn.

CLAS has collected ~$5\cdot 10^5$ $p\eta^{\prime}$ events
in photoproduction. Production of $\eta^{\prime}$ has 
also been observed in electron scattering for the first time with CLAS. 
This channel may provide a new tool in the search for missing states.
The quark model predicts two resonances in this mass range 
with significant coupling to the $N\eta^{\prime}$ channel \cite{capstick}. 

$K\Lambda$ or $K\Sigma$ production may yet be another source of information on resonant
 states. Previous data show some evidence for resonance production in these 
 channels \cite{saphir}.
New data with much higher statistics are being accumulated with the CLAS
detector, both in photo- and electroproduction. 
Strangeness production could open up yet another window for light quark baryon spectroscopy,
which was not available in the past.

\begin{figure}[htbp]
\begin{minipage}{0.45\textwidth}
\epsfysize=7.5cm
\epsfxsize=7.5cm
\epsfbox{fig8_e99.epsi}
\hsize=7.5cm
\caption{\small  Electroproduction of $\omega$ mesons for different W bins. The deviation 
of the $\cos\theta$ -distribution from a smooth fall-off for the low W bin suggests significant 
s-channel resonance production.}
\end{minipage}
\begin{minipage}{0.45\textwidth}
\epsfysize=8.0cm
\epsfxsize=7.5cm
\hspace{1.0cm}
\epsfbox{fig9_e99.epsi}
\hspace{1.0cm}
\hsize=8.cm
\caption{\small Ratio of resonance excitations as observed and predicted from 
deep inelastic processes using quark-hadron duality.\cite{keppel}}
\end{minipage}
\end{figure}

\section{\bf Local Duality - Connecting Constituent Quarks and Valence Quarks}

I began my talk by expressing the expectation that we may eventually 
arrive at a unified description of hadronic structure from small to 
large distances. If such description is possible then there should 
be obvious connections in the data between these regimes. 
Such strong connections have indeed been observed by 
Bloom and Gilman \cite{blogil}. They noted that the scaling curves from 
the deep inelastic cross section also describe the average 
inclusive cross sections in the resonance region if a scaling variable is chosen 
that takes into account target mass effects. Until recently, this intriguing
observation was little utilized. 
A new inclusive ep scattering experiment at JLAB \cite{keppel} helped rekindle
the interest in this aspect of hadron physics. 
Remarkably, elastic form factors or resonance excitations of the nucleon
can be predicted approximately from inclusive deep 
inelastic scattering data.  
Figure 9 shows the ratio of measured integrals over resonance 
regions, and predictions using deep inelastic data only. The agreement
is surprisingly good, though not perfect, indicating that the concept 
of duality likely is a non-trivial consequence of the underlying dynamics.

How can this success be explained in terms of the underlying degrees of 
freedom - elementary partons, and constituent quarks, respectively. Part of the answer 
is shown in Figure 10, where a large number of resonance data sets with different 
$Q^2$ are shown together with the evolution curves from deep inelastic 
scattering. The deep inelastic curves fail to reproduce the resonance 
data at small $\xi$, while an evolution using valence quarks only has the same 
small $\xi$ behavior.  A new fit to the data (labelled Jlab fit) reproduces the 
the $xF_3(x)$ structure function determined from the difference of neutrino and 
anti-neutrino scattering (Figure 11) This quantity only contains valence quarks. 
The agreement suggests that the constituent quark distribution in the resonance 
region has an $\xi$ dependence very similar to the distribution of elementary 
valence quarks in the deep inelastic region. It remains to be seen if this intriguing
observation can be translated into the development of new model approaches to 
resonance physics.

\begin{figure}[htbp]
\begin{minipage}{0.45\textwidth}
\epsfysize=8.truecm
\epsfbox{fig10_e99.epsi}
\hsize=8.0cm
\caption{\small Compilation of resonance data at different $Q^2$. The curves are from
the evolution of deep inelastic data, with the exception of the curve labelled `JLab fit'
which represents a new fit to the JLab data. }
\end{minipage}
\begin{minipage}{0.45\textwidth}
\epsfysize=8.truecm
\epsfbox{fig11_e99.epsi}
\hsize=8.0cm
\caption{\small The JLab fit to the $F_2$ data from Figure 10 shown together with the 
$xF_3$ structure function obtained from neutrino and anti-neutrino  
data. The latter represent the valence quark distribution in the nucleon.}
\end{minipage}
\end{figure}

\noindent
In the following I will discuss recent results from experiments on $^2H$
and $^3He$. 

\section{Elastic Formfactors of the Deuteron}

In the same way that elastic electron scattering on protons and neutrons 
reveals their intrinsic charge and current distributions, so does elastic 
electron-deuteron scattering. Since the deuteron has spin 1, the elastic
response functions contain 3 electromagnetic form factors $G_C$, $G_Q$, and $G_M$. 
On the one hand, the interest
in studying these form factors is to obtain a 
complete set of 
measurements. This involves measurement of at least one polarization 
observable ($T_{20}$). On the other hand, there is significant interest in 
the high $Q^2$ behavior. There we probe the short distance behavior of 
the nucleon-nucleon interaction.
A large variety of models have been developed to describe the form factors
for a wide range in distance scale, from hadronic models that 
include nucleons, pion, isobars, and exchange currents, descriptions within 
the quark exchange picture, to descriptions within the framework of 
perturbative QCD.

\subsection{Unpolarized elastic response functions in $eD \rightarrow eD$}

The unpolarized elastic eD scattering cross section contains the 
two response functions A, B: 

$$ {d\sigma \over d\Omega} = \sigma_M [A(Q^2) + B(Q^2)tan^2({\theta \over 2})]~~,$$    

\noindent 
where 
$$A(Q^2) = G^2_C(Q^2) + {8\over 9}\tau^2 G^2_Q(Q^2) + {2\over 3}\tau G^2_M(Q^2)$$
$$B(Q^2) = {4\over 3} \tau (1+\tau)G^2_M(Q^2);~~~~~~\tau = {Q^2\over 4M^2}$$  

\noindent
Unpolarized electron scattering allows determination of 
the magnetic form factor by 
measuring the scattered electron at backward angles. A separation of $G_C$ and $G_Q$ 
is not possible. One can only separate the response functions $A(Q^2)$ and $B(Q^2)$
by measuring the elastic cross section at fixed $Q^2$ and different scattering angles
(Rosenbluth separation). An experiment in JLab Hall A (E-91-026) measured this process
in a coincidence setup, where both the scattered electron and recoil deuteron were 
detected in two high resolution spectrometers. The results for $A(Q^2)$ are 
shown in Figure 13. 

The data are approximately described by modern hadronic models. Even
the approach to scaling may therefore be understood within these models. 
It is therefore not obvious that quark-gluon degrees of freedom have to be invoked 
to describe the data even at the highest momentum transfers.

\begin{figure}[htbp]
\epsfysize=9.0cm
\epsfxsize=8.5cm
\epsfbox{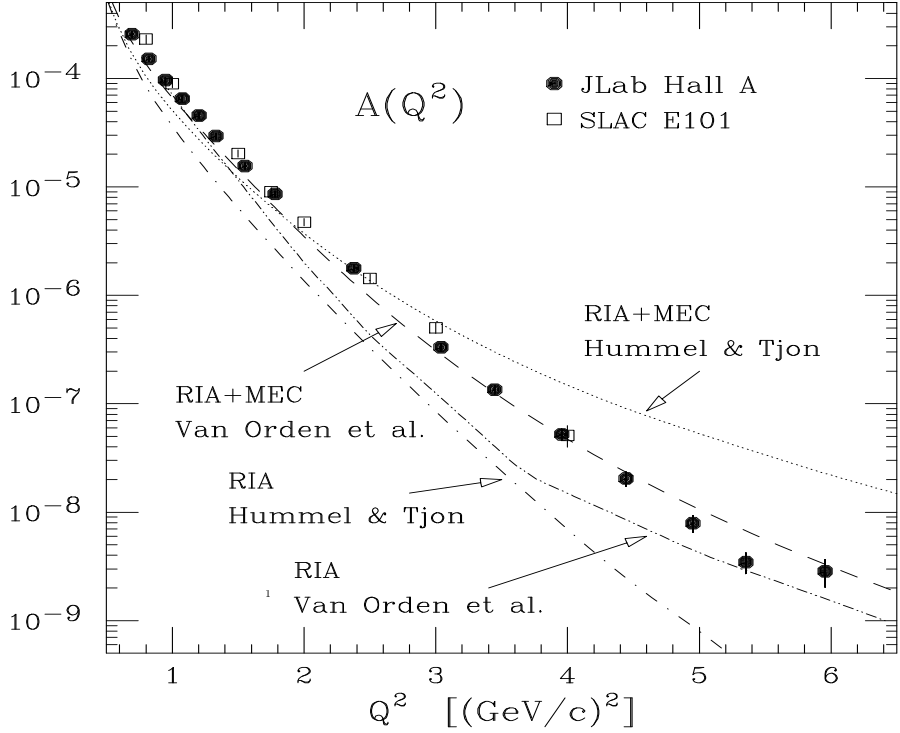}

\vspace{0.95cm}
\hsize=13.5cm
\caption{\small The electric response function $A(Q^2)$ measured 
in $eD\rightarrow eD$ scattering. The JLab data extend the $Q^2$ range 
from previous SLAC data.}
\end{figure}

\subsection{Tensor Polarization in $eD \rightarrow eD$}  

A separation of the charge and quadrupole form factors of the deuteron  requires 
a polarization experiment, in addition to the unpolarized measurement. 
A measurement of the tensor polarization $t_{20}$ is particularly 
suited to accomplish this. This tensor polarization component can be expressed in 
terms of the electromagnetic form factors:

$$t_{20} = -2 {{\tau G_Q(\tau G_Q + 3G_C)} 
\over {G^2_C + {8\over 9}\tau^2 G^2_Q}}$$  

These experiments require a measurement of the deuteron recoil polarization in a 
second scattering experiment using a suitable analyzing reaction. Previous experiments
of this type have been carried out at lower energy accelerators, covering the 
lower $Q^2$ range. The JLab experiment was carried out in Hall C, 
using a high power deuterium cryogenic target, and a new deuteron
magnetic spectrometer to analyze the kinematics and deuteron polarization. A liquid 
hydrogen target was used for the second scattering experiment, which was needed 
to analyse the deuteron polarization\cite{kox}. The results for 
 $t_{20}$ are shown in Figure 14 \cite{real,hafadi}. Using the known response function $A(Q^2)$
the deuteron charge form factor $G_C(Q^2)$ can be separated (Figure 15). The charge 
form factor shows a zero crossing at $Q^2 = 0.7~GeV^2$, and remains
negative over the complete large $Q^2$ range.  

\noindent
Hadronic models describe the data over the entire range in momentum transfer.

\begin{figure}[htbp]
\begin{minipage}{0.4\textwidth}
\epsfysize=8.0cm
\epsfxsize=8.0cm
\epsfbox{fig13_e99.epsi}
\hsize=8.0cm
\caption{\small The tensor polarizations $t_{20}$, $t_{21}$ and $t_{22}$  
measured in $eD\rightarrow eD$. 
The deuteron polarization was measured in $Dp \rightarrow ppn$ scattering.}
\end{minipage}
\begin{minipage}{0.4\textwidth}
\epsfysize=8.0cm
\epsfxsize=8.0cm
\epsfbox{fig14_e99.epsi}
\hsize=8.0cm
\hspace{1cm}
\caption{\small The deuteron charge and quadruole form factors 
as extracted from the $t_{20}$ measurement and the known $A(Q^2)$ response function.}
\end{minipage}
\end{figure}

\subsection{Deuteron photo-disintegration at high momentum transfer}

The deuteron is an ideal laboratory to study where the traditional Yukawan 
picture of the nucleus may break down, and the quark picture may provide a more
effective description. The deuteron as the simplest nucleus permits
exact hadronic calculations. Experimentally, one can give 
a large momentum transfer to the constituents, and thus study the approach 
to scaling at modest energies. 

One of the indications for the relevance of quark constituents in the interaction
is scaling of the differential cross section according to the number
of constituents involved in the interaction. Constituent counting rules 
predict that the energy dependence for the two-body reaction $\gamma d \rightarrow 
n p$ should scale like:
$$ {d\sigma \over dt} = {h(\theta_{cm}) \over s^{n-2}}~~,$$
where n is the number of elementary fields in the initial and final states, and 
n-2 = 11 for the   $\gamma d \rightarrow n p$. 
While scaling has been observed at center-of-mass angles near 90$^o$ for 
photon energies as low as 1 GeV and up to the maximum energies of 4 GeV (Figure 16), 
no scaling is observed for smaller $\theta_{cm}$ angles \cite{holt1}.

 New models have been developed that give a more realistic 
 description of the process, and predict the parameter n to be angle-dependent,
  e.g. the 
 Regge gluon-string model 
 \cite{desanctis}, and quark exchange 
 models  where the number of constituent involved in the 
 reaction is smaller than in the maximal model ("constituent scaling")
 which involves all constituents.  These models indeed provide a better 
 description of the reaction over a larger kinematical range (Figure 17).

New data have been taken to extend the kinematic range up to 
5.5 GeV photon energies to see whether scaling persists for the 
$90^o$ kinematics, and if scaling is approached at different angles.
 
\subsection{Polarization asymmetries on $^3He$}

$^3He$ has emerged as an attractive target material for polarized neutrons. 
It is the closest any nucleus comes 
to a pure polarized neutron target and is a simple enough nucleus, so that 
corrections to this naive picture can be calculated with some confidence. 
At low momentum transfer, corrections appear to be large for some reactions.  

An experiment in JLAB Hall A measured quasi-elastic electron scattering off $^3He$ 
in an effort to get information on the magnetic form factor of the neutron, and
to study asymmetries in the breakup region at small excitation energies
\cite{gao}. 

Figure 18 shows preliminary data for the sensitivity of the measured asymmetry
to the neutron magnetic form factor. The model dependency of final state 
corrections seems small enough to allow extraction of the quantity of 
interest.

\begin{figure}[htbp]
\begin{minipage}{0.45\textwidth}
\epsfysize=8.0cm
\epsfxsize=7.5cm
\epsfbox{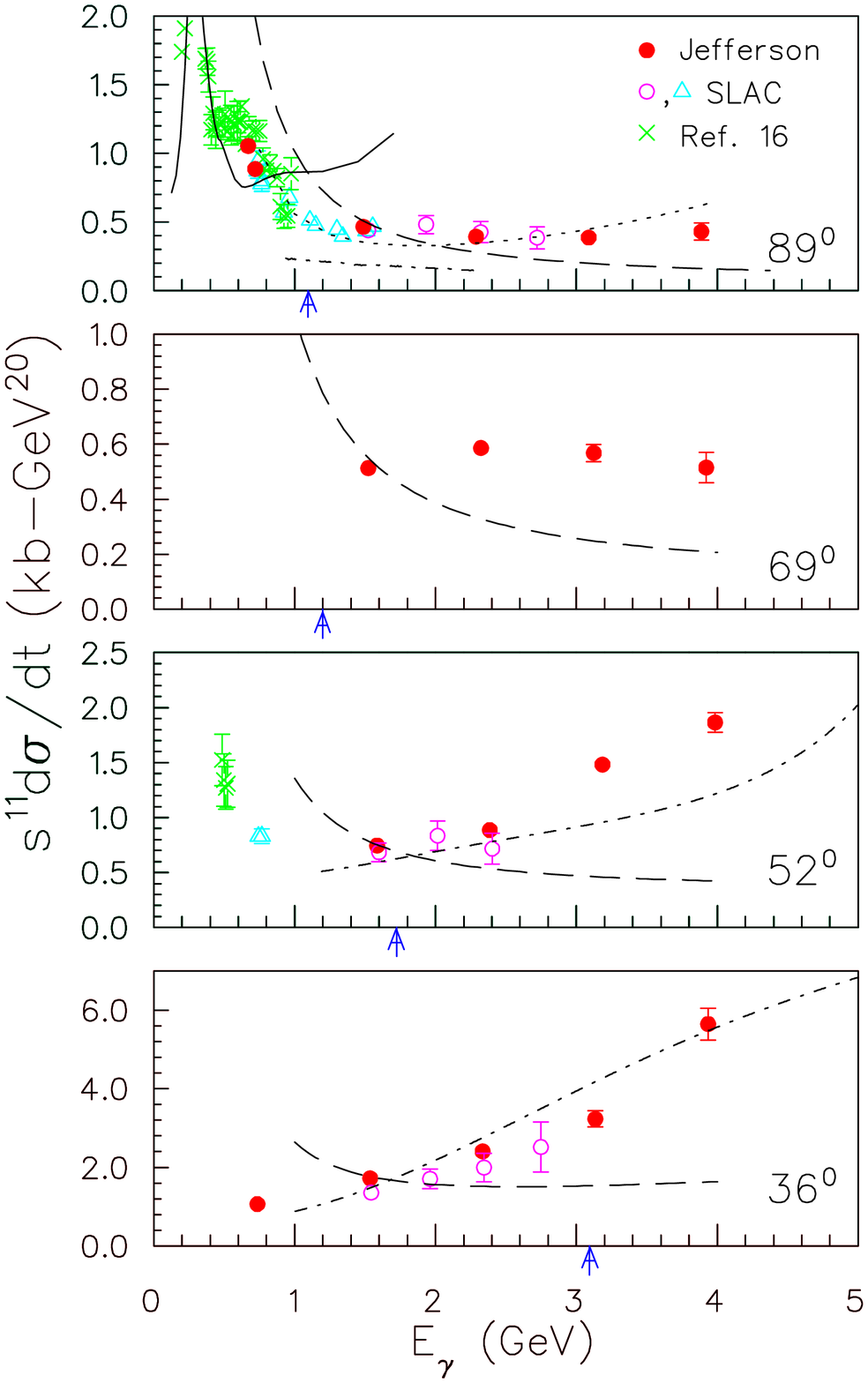}
\hsize=8.0cm
\caption{\small The cross section for $\gamma d \rightarrow n p$ multiplied by 
the predicted dimensional scaling function $s^{11}$. Scaling is not observed at 
small angles, where the quark exchange model gives a better representation 
of the data.}
\end{minipage} 
\begin{minipage}{0.45\textwidth}
\epsfysize=8.cm
\epsfxsize=8.cm
\epsfbox{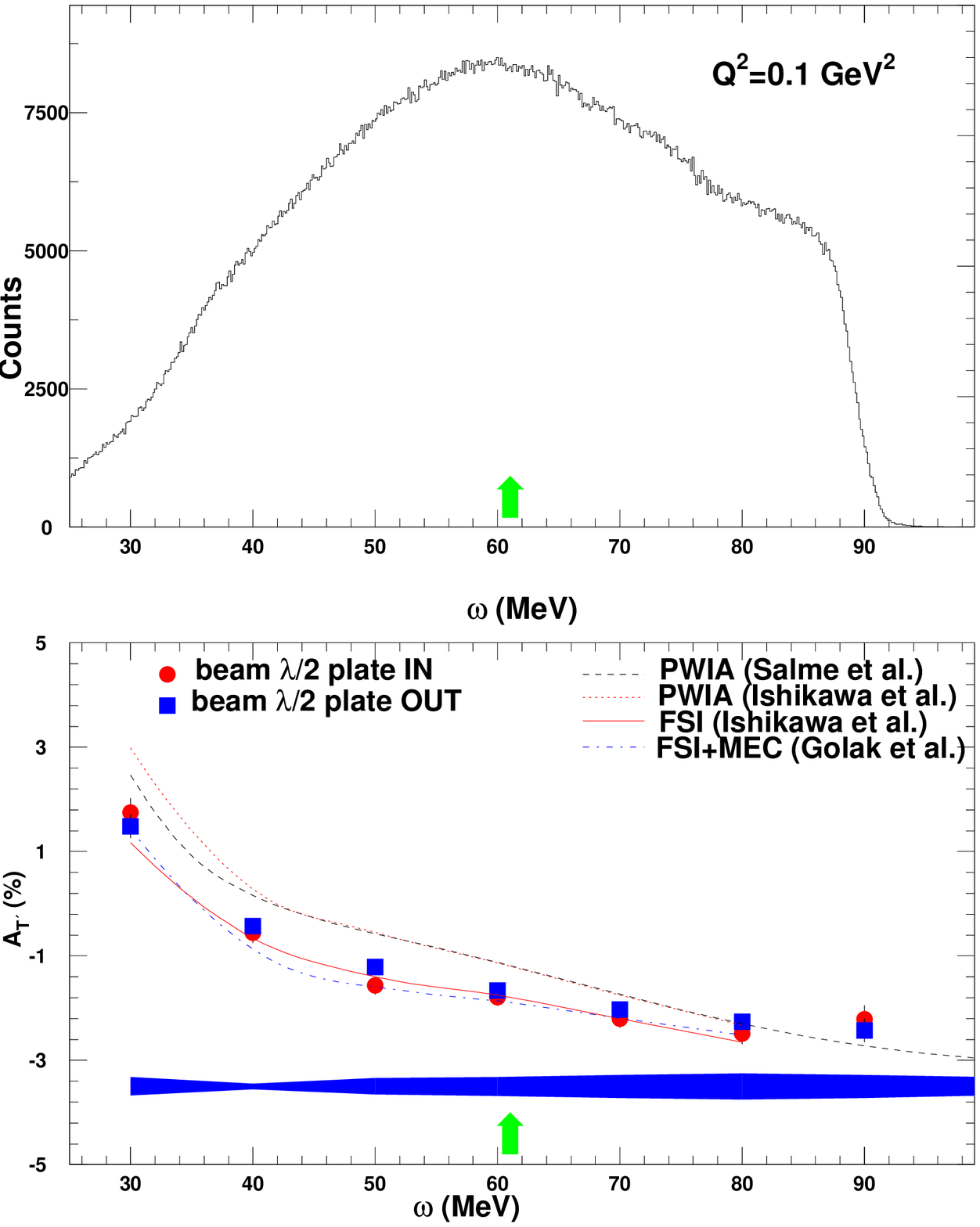}
\hsize=7.5cm
\caption{\small Preliminary results for JLAB experiment E95-001 to measure the 
magnetic form factor of the neutron. The experimental asymmetry is shown 
measured in scattering of polarized electrons from a polarized $^3He$ target.}
\end{minipage}
\end{figure}

\section{Outlook}

The ongoing experimental effort at Jefferson Lab will provide the community 
with a wealth of 
data in the first decade of the next millennium to address many open problems 
in hadronic structure at intermediate distances. 
{\it The experimental effort must be accompanied by a significant theoretical effort 
to translate this into real progress in our understanding of the
complex regime of strong interaction physics}. One area, where a fundamental 
description may be within reach, is the evolution of the nucleon spin structure 
from small to large distances.

New instrumentation will become available, e.g. the $G^o$ experiment 
at JLAB, allowing a broad program in parity violation to 
study strangeness form factors in electron scattering in a large kinematic range.
    
Moreover, there are new opportunities on the horizon. Recently, it was 
shown\cite{ji,radyu} that in exclusive processes 
the soft (nonperturbative) part and the hard (perturbative) parts factorize 
for longitudinal photons at sufficiently high $Q^2$. 
A  new set of  ``skewed parton distributions'' can then be measured which are 
generalizations of the inclusive structure functions measured in
 deep inelastic scattering. 
For example, low-t $\rho$ production probes the unpolarized parton 
distributions, while pion production probes the polarized structure functions. 
Experiments to study these new parton distributions need to have
sufficient energy transfer and momentum transfer to reach the pQCD regime, 
high luminosity to measure the small exclusive cross sections, and
good resolution to isolate exclusive reactions.

This new area of research may become a new frontier of electromagnetic 
physics well into the next century. 

To accommodate new physics requirements, an energy upgrade in the 10-12 GeV 
range has 
been proposed for the CEBAF machine at JLAB. 
This upgrade will be accompanied by the construction of a new experimental 
hall for tagged photon experiments with a 4$\pi$ solenoid detector to study 
exotic meson spectroscopy, and production of other heavy mesons.
Existing spectrometers in Hall C will be upgraded to reach higher momenta 
and improvements of CLAS will allow it to cope with higher multiplicities.

This will give us access to kinematics where copious hybrid meson 
production is expected, higher momentum transfer 
can be reached for form factor measurements, and we may begin to map out 
the new generalized parton distributions.

\end{document}